# Ge doping of GaN beyond the Mott transition

A. Ajay[1,2], J. Schörmann[3], M. Jiménez-Rodriguez[1,2,4], C. B. Lim[1,2], F. Walther[3], M. Rohnke[6], I. Mouton[1,2], L. Amichi[1,2], C. Bougerol[1,5], M. I. Den Hertog[1,5], M. Eickhoff[3] and E. Monroy[1,2]

[1] Université Grenoble-Alpes, 38000 Grenoble, France.
[2] CEA-Grenoble, INAC, 17 av. des Martyrs, 38000 Grenoble, France.
[3] I. Physikalisches Institut, Justus-Liebig-Universität Gießen and Center for Materials Science, Heinrich-Buff-Ring 16, 35392 Gießen, Germany.
[4] GRIFO, Dept. Electrónica, Universidad de Alcalá, 28871 Alcalá de Henares, Madrid, Spain.
[5] CNRS-Institute Néel, 25 av. des Martyrs, 38000 Grenoble, France.
[6] Physikalisch-Chemisches Institut and Center for Materials Science, Justus-Liebig-Universität Gießen, Heinrich-Buff-Ring 17, 35392 Gießen, Germany

**Abstract**

We present a study of germanium as n-type dopant in wurtzite GaN films grown by plasma-assisted molecular-beam epitaxy, reaching carrier concentrations of up to $6.7 \times 10^{20}$ cm$^{-3}$ at 300 K, well beyond the Mott density. The Ge concentration and free carrier density were found to scale linearly with the Ge flux in the studied range. All the GaN:Ge layers present smooth surface morphology with atomic terraces, without trace of pits or cracks, and the mosaicity of the samples has no noticeable dependence on the Ge concentration. The variation of the GaN:Ge band gap with the carrier concentration is consistent with theoretical calculations of the band gap renormalization due to electron-electron and electron-ion interaction, and Burstein-Moss effect.

# 1. Introduction

Silicon has been the preferred n-type dopant for wurtzite GaN, even though it contributes to edge type dislocation climb, leading to an increase in tensile stress [1–3]. Tensile strain is especially troublesome when growing GaN on silicon substrates, where the mismatch of thermal expansion coefficients requires careful strain engineering to avoid cracking [4]. Furthermore, doping with Si in excess of $10^{19}$ cm$^{-3}$, i.e. above the Mott transition in GaN, is reported to cause surface roughening and eventually crack propagation [2]. In nanowire structures, currently under investigation for a new generation of light emitting diodes, the radial distribution of Si is inhomogeneous, with a tendency to migrate towards the sidewalls [5], and high doping levels tend to degrade the nanowire morphology [6].

In response, Ge was recently reintroduced as a highly favorable n-type dopant in GaN, with the possibility for higher doping concentrations [2,3]. Ge, like Si, is a shallow donor in GaN, with a theoretical activation energy of 31.1 meV [7]. The ionic radius of a Ge atom is similar to that of Ga and the metal-nitrogen bond length changes by only 1.4% with Ge, compared to 5.5% with Si [8]. Hence Ge can occupy the Ga lattice site causing far less lattice distortion than other dopants like Si and O. Also like Si, the DX- state of Ge is unstable and does not affect doping efficiency [8]. From both perspectives, fundamental and applied, it is important to perform extensive studies on Ge as a dopant in GaN.

A few studies regarding successful Ge doping of GaN using hydride vapor phase epitaxy (HVPE) [9,10] and metalorganic vapor phase epitaxy (MOVPE) [11–14] exist, and recent results on Si(111) substrates show the absence of additional tensile stress [3,13]. Using plasma-assisted molecular-beam epitaxy (PAMBE), free carrier concentration up to $4\times10^{20}$ cm$^{-3}$ were reported in thin films [15,16], and nanowires doped with Ge were also demonstrated to achieve significant dopant levels and metallic conductivity [17,18]. In GaN/AlN nanowire heterostructures, the introduction of Ge doping has led to the demonstration of screening of the internal electric field [19,20], and to the observation of intersubband transitions [21]. However, there is still a large void regarding the effect of Ge doping on the PAMBE growth kinetics and properties of planar GaN:Ge.

In this work, we expand on the existing literature by analyzing the effect of Ge on the PAMBE growth kinetics of GaN complimented with a comprehensive analysis on the optical and electrical properties of GaN:Ge which vary in accordance with the variation in carrier concentration. Carrier concentrations are quantified using Hall effect measurements and infrared spectroscopy. We demonstrate carrier concentrations up to $6.7 \times 10^{20}$ cm$^{-3}$ (i.e. 1.5% atomic incorporation), well above the Mott transition, which do not perturb the growth process, the layer morphology or its structural quality. The GaN:Ge photoluminescence (PL), and band gap evolution with the doping level are consistent with theoretical calculations of the band gap renormalization and Burstein-Moss effect.

## 2. Experimental

GaN thin films with a thickness of 675 nm were grown by PAMBE on 1-µm-thick AlN-on-sapphire templates. The growth rate was 0.5 ML/s (≈450 nm/h) and the growth temperature was $T_S$ = 720°C for all the experiments. The epitaxial process was monitored by reflection high energy electron diffraction (RHEED), and the growth temperature was verified by measurement of the Ga desorption time from the GaN(0001) surface [22]. Different samples were obtained by doping with different amounts of Ge by varying the Ge cell temperature, $T_{Ge}$, from 720°C to 1000°C (see list of samples in table 1). The sample morphology was analyzed by atomic force microscopy (AFM) using a Veeco Dimensions 3100 system operated in tapping mode. The structural quality of the layers was further studied by high-resolution x-ray diffraction (HRXRD) using a PANalytical X'Pert PRO MRD in triple-axis configuration. Room-temperature and temperature-dependent Hall effect measurements were performed to characterize carrier concentration and activation energy in the Van der Pauw geometry on samples contacted by e-beam evaporation of Ti/Al (30/150 nm).

Fourier transform infrared spectroscopy (FTIR) was used to identify bulk plasma oscillations using a Bruker V70v spectrometer. Reflectivity measurements were performed with p-polarized light at 70° incidence. The resulting spectrum was corrected by the system response by dividing it by the reflectivity measurement of an Au film using the same experimental conditions. The experiment was repeated for other incident angles to identify interference oscillations. Reflections using s-polarized light showed interference fringes associated to the thickness of the GaN layer. Time-of-flight secondary ion mass spectroscopy (ToF-SIMS) was performed using a TOF-SIMS 5 system (ION-TOF GmbH) equipped with a 25 keV Bi-cluster primary ion gun and a 2 keV Cs-sputter gun.

Photoluminescence (PL) spectra were obtained by excitation with a continuous-wave solid-state laser ($\lambda$ = 244 nm), with an excitation power around 100 µW focused on a spot with a diameter of 100 µm. The emission from the sample was collected by a Jobin Yvon HR460 monochromator equipped with an ultraviolet-enhanced CCD camera. The evolution of the band gap at room temperature was studied by optical transmission under normal incidence, exciting with a 450 W Xe-arc lamp coupled to a Gemini-180 double monochromator in the range of 320 nm to 420 nm.

## 3. Results and discussion

Non-intentionally doped (n.i.d.) Ga-face GaN is synthesized by PAMBE under slightly Ga-rich conditions, with a 2-ML-thick self-regulated Ga adlayer on the growing surface [23]. The Ga adlayer is known to be sensitive to the presence of dopants. Particularly, silicon does not introduce any perturbation in the Ga kinetics; whereas segregation of Mg inhibits the formation of the Ga adlayer [24–26] and segregation of Mg on the growing surface had significantly reduced the growth window in terms of Ga flux [25]. To analyze the effect of Ge on the adatom kinetics, we have studied the Ga desorption during a growth interruption after the deposition of 7 nm Ge-doped GaN for various Ge fluxes. The value of the Ga flux is approximately $\approx 0.7$ ML/s, chosen so that the 2-ML Ga adlayer is dynamically stable during the growth of n.i.d. GaN, and an increase of Ga flux by 7% leads to the accumulation of Ga on the growing surface. After each measurement of Ga desorption from GaN:Ge, we deposit 7 nm of n.i.d GaN and record the Ga desorption from the undoped surface as a reference. Figure 1 presents the RHEED intensity transient generated by the Ga desorption after the growth of n.i.d GaN and Ge-doped GaN for Ge cell temperatures $T_{Ge}$ = 800°C, 900°C, 950°C and 1000°C. The growth was interrupted at time $t$ = 0 by closing the Ge, Ga and N cells simultaneously. The RHEED intensity transients remain unaltered with the presence of a Ge flux, i.e. Ge does not perturb the Ga kinetics on the GaN(0001) growth front.

Following this information, Ge-doped GaN thin films with a thickness of 675 nm were grown in the 2-ML Ga-adlayer regime on 1-µm-thick AlN-on-sapphire templates. After the growth, the surface morphology was analyzed by AFM with the results presented in figure 2. Regardless of the doping level, all the samples present monoatomic terraces and hexagonally-shaped hillocks characteristic of PAMBE-grown GaN, without observation of cracks or pits. The root-mean-square (rms) roughness measured in

5×5 µm² images is 0.9±0.3 nm for all the samples. These results demonstrate that there is no significant effect of Ge on the surface morphology, in contradiction to SEM observations reported in ref [15].

The structural quality was further examined by HRXRD. From ω-2θ scans of the (0002) GaN reflection, the average strain of the c-lattice parameter of GaN was estimated to $\varepsilon_{zz} = 0.22 \pm 0.07\%$, without any clear trend as a function of the Ge concentration (see strain of samples in table 1). This tensile strain along c is the result of the compressive in-plane stress imposed by the AlN substrate, in good agreement with previous studies of the plastic relaxation of GaN on AlN when growing by PAMBE in the 2-ML Ga-adlayer regime [27]. The full width at half maximum (FWHM) of the ω-scan of the (0002) reflection of GaN:Ge is compared to the n.i.d reference sample in table 1. For all the layers, the FWHM remains 190±110 arcsec without any systematic influence of the Ge incorporation.

Results of the electrical characterization by Hall Effect at 300 K are given in table 1. Free carrier concentrations of up to $n$ = 6.7×10²⁰ cm⁻³ are obtained for the sample grown with the highest Ge cell temperature. As illustrated in figure 3(a), $n$ scales exponentially with the Ge cell temperature, following $n \propto \exp(-E_{Ge}/k_B T_{Ge})$, where $E_{Ge} = 3.58$ eV is the thermal evaporation energy of Ge, and $k_B$ is the Boltzmann constant. This result points to a Ge concentration of at least 6.7×10²⁰ cm⁻³, which corresponds to a Ge mole fraction higher than 1.5% in the layer. The resistivity at 300 K (also in table 1) decreases over two orders of magnitude when increasing the free carrier concentration from 7.9×10¹⁷ cm⁻³ to 6.7×10²⁰ cm⁻³, reaching a lowest value of 6.90×10⁻⁴ Ωcm (sample I).

The carrier concentration of the most heavily doped samples (samples F to I) was also estimated from the plasma frequency ($\omega_p$) using mid-infrared reflectivity measurements with p-polarized light, in figure 4. According to the Drude-Lorentz model, the plasma frequency of the free carrier plasma is given by

$$\omega_p = \frac{ne^2}{m_e^* \varepsilon_s \varepsilon_0} \quad (1)$$

where $e$ is the elementary charge, $\varepsilon_0$ is the vacuum permittivity, $\varepsilon_s = 9.38$ is the static dielectric constant of GaN, and $m_e^* = 0.231 m_0$ is the electron effective mass in GaN. The estimated free carrier concentrations are illustrated in figure 3(a) and are similar to the data obtained from Hall effect measurements.

For further analysis of the incorporation of Ge, all samples were analyzed by ToF-SIMS. The depth profile of GaN:Ge sample H and the n.i.d reference sample X are depicted in figures 3(c, d). A uniform Ge signal throughout the thickness of the GaN layers is found. Oxygen and trace amounts of Si and C, seen in all samples including the reference, have no specific dependence on $T_{Ge}$ and can be concluded to be almost constant from sample to sample. It should be noted that unintentional n-type dopants (Si and O) are two and three orders of magnitude lower than Ge. The carrier concentration extracted from Hall Effect measurements at room temperature scales linearly with the Ge signal obtained from ToF-SIMS, as shown in figure 3(b), confirming that the variation in carrier concentration is indeed due to Ge incorporation. We have made an effort to determine our error bar in the quantification of Ge, which is described in the annex.

Figures 5(a) and (b) describe the variation of carrier concentration and resistivity with temperature in samples with doping levels up to the density of the Mott transition (~1-1.5×10$^{19}$ cm$^{-3}$ in GaN at room temperature) [28,29]. The activation energies $E_a$ extracted from figure 5(a) decrease from 19.5 meV for the lowest doped sample A, to 12.4 meV for sample B, and 9.7 meV for sample C, in accordance with the decrease in average distance between the impurity atoms [30]. Sample D presents metallic behavior.

Thus, the effective activation energy can be described by $E_a = E_I - \alpha(n)^{1/3}$, where $E_I$ is the activation energy of a single isolated Ge impurity atom in GaN (theoretically estimated at $E_I$ = 31.1 meV by Wang and Chen [7]), and $\alpha$ is a proportionality constant. The value of $\alpha$ determined by the empirical fit is (1.6±0.3)×10$^{-5}$ meVcm, close to the $\alpha$ = (2.1±0.2)×10$^{-5}$ meVcm reported for Si donors in GaN [31]. We do not take into account the unintentional dopants like O and Si which are two and three orders of magnitude lower than Ge, as discussed above.

The PL emission characteristics were analyzed, and spectra recorded at low temperature ($T$ = 5 K) and room temperature are displayed in figures 6(a) and (b), respectively. The PL spectra at 5 K of n.i.d GaN (sample X) show near-band-edge excitonic emission around 3.515 eV, blue-shifted with respect to the theoretical value (in the range of 3.470 eV in bulk GaN) due to the in-plane compressive strain induced by the AlN substrate. Regardless of the temperature, as the Ge concentration increases, for samples A to C, a redshift is noticed consistent with band gap renormalization (BGR) [32] due to electron-electron and electron-ion interaction. Further increase of carrier concentration (samples D to I) causes a blue-shift of the emission due to the Burstein-Moss effect (BME) [33], i.e. the lower energy states in the conduction band become significantly filled, and the Fermi level lies inside the conduction band. The emission spectra of samples D to I are systematically broadened with the increase in carrier concentration. Their shape corresponds to the Kane density of states for the conduction band multiplied by the Fermi-Dirac distribution. Our observations are similar to the description in ref. [34], which used GaN doped with Si in the range of 8.7×10$^{-17}$ cm$^{-3}$ to 1.4×10$^{-19}$ cm$^{-3}$ synthesized using HVPE, and Ge for dopant concentrations in the range of 3.4×10$^{-19}$ cm$^{-3}$ to 1.6×10$^{-20}$ cm$^{-3}$, synthesized using MOVPE.

The evolution of the band gap at room temperature was studied by optical transmission in the range of 320 nm to 420 nm. Measurements were performed under normal incidence, exciting with a 450 W Xe-arc lamp coupled to a Gemini-180 double monochromator. The band gap energy ($E_G$), determined from a Tauc's plot as displayed in the inset of figure 7, is provided in table 1. Similar to the low-temperature PL emission, the value of $E_G$ first redshifts (samples A to C) and then blue-shifts (samples D to I) with increasing carrier concentration, as illustrated in figure 7. In addition to the spectral shift, the increase in doping concentration also affects the slope of the absorption edge that can be described by introduction of an Urbach energy $\Delta E_{Urb}$ in the expression for the optical absorption: $\alpha(\lambda) = \alpha_0 \exp[(hc/\lambda - E_0)/\Delta E_{Urb}]$, where $hc/\lambda$ is the photon energy, $\alpha_0$ and $E_0$ are material-dependent fitting parameters. The values of $\Delta E_{Urb}$ extracted from transmission measurements are summarized in table 1, showing a monotonous increase with the carrier concentration, as expected.

The shift of the fundamental band gap with the free carrier concentration is given by the superposition of BGR and BME, i.e. $\Delta E = \Delta E_{BMS} + \Delta E_{BGR}$. We estimate individual contributions analytically similar to ref. [35]. The BGR shift itself has two contributions due to electron-electron and electron-ion interactions ($\Delta E_{ee}$ and $\Delta E_{ei}$, respectively), which can be approximated as

$$\Delta E_{ee} = -\frac{e^2 k_F}{2\pi^2 \varepsilon_s \varepsilon_0} - \frac{e^2 k_{TF}}{8\pi \varepsilon_s \varepsilon_0}\left[1 - \frac{4}{\pi}\arctan\left(\frac{k_F}{k_{TF}}\right)\right] \qquad (2)$$

$$\Delta E_{ei} = -\frac{e^2 n}{\varepsilon_s \varepsilon_0 a_{B^-}^* k_{TF}^3} \qquad (3)$$

where $k_F = (3\pi^2 n)^{1/3}$ is the Fermi vector, $k_{TF} = 2\sqrt{k_F/(\pi a_{B^-}^*)}$ is the inverse Thomas-Fermi screening length, and $a_{B^-}^* = 4\pi\varepsilon_s\varepsilon_0\hbar^2/(m_e^* e^2)$ is the effective Bohr radius of the electron, with $\hbar$ being the reduced Planck constant.

On the other hand, the shift induced by the BME follows the equation

$$\Delta E_{BMS} = \frac{\hbar^2 k_F^2}{2\mu^*} \qquad (4)$$

where $\mu^*$ is the reduced effective mass.

The contribution from excitonic effects, influenced by the carrier concentration, is weak at high carrier densities and is hence neglected here. A plot of $\Delta E$ is displayed in figure 7, showing good agreement with the band gap values obtained from Tauc's plot, which confirms that the trend in optical properties is in correlation with the trend in carrier concentration.

## 4. Conclusions

In conclusion, the effect of Ge doping in wurtzite GaN grown by PAMBE has been studied systematically. It was verified that the presence of a Ge flux does not perturb the Ga kinetics during the growth of GaN, even for $T_{Ge}$ = 1000°C. We have synthesized GaN:Ge with free carrier concentrations in the range of 7.9×10$^{17}$ cm$^{-3}$ to 6.7×10$^{20}$ cm$^{-3}$. We observed that the free carrier concentration scales linearly with the Ge incorporation. Regardless of the Ge concentration, all samples kept indistinguishable surface morphology with atomic terraces and a roughness in the 0.9±0.3 nm range. The FWHM of the x-ray rocking curve in the 190±110 arcsec range does not show any trend as a function of the doping level, showing that the mosaicity of the samples has no noticeable dependence on Ge doping in the range under study. Optical studies demonstrate variations of the GaN:Ge band gap consistent with theoretical calculations of the band gap renormalization due to electron-electron and electron-ion interaction and Burstein-Moss effect. We were able to correlate optical properties to the carrier concentration which arises from the Ge incorporation, In view of these results, the use of Ge appears as a satisfactory alternative to Si for n-type doping of GaN grown by PAMBE, particularly above the Mott density, where structural defects appear associated to Si incorporation.

**Acknowledgements.** The authors acknowledge technical support by D. Boilot, Y. Genuist and Y. Curé, and thank B. Gayral for fruitful discussions. We also thank M. Lafossas and A. Dussaigne from CEA-LETI for validating our Hall measurements. This work is supported by the EU ERC-StG "TeraGaN" (#278428) project, and by the French National Research Agency via the GaNEX program (ANR-11-LABX-0014).

## Annex: Error bar in the quantification of Ge concentration

To give a better idea of the error bars associated to the quantification of Ge, we summarize here the efforts that were made to validate our results. On one hand, measurements of carrier concentration were performed by Hall effect in two different setups, and by infrared reflectometry. On the other hand, Ge atom density measurements were performed using ToF-SIMS, standard SIMS, and atom probe tomography (APT).

Hall effect measurements were made in the Van der Paw geometry on samples contacted by e-beam evaporation of Ti/Al (30/150 nm). Measurements were performed at constant current and varying the magnetic field in the range of −0.8 to 0.8 T. For each sample, several measurements were made with various input currents in the range of 0.1 to 1 mA. Average values are presented in table 2 ($n_A$), where the error bars correspond to the error in the determination of the layer thickness (below 10%). The standard deviation of the Hall factor measurements for the same sample measured at different current values was always lower than 3%.

To validate these Hall effect results, one of the contacted samples has been measured in another laboratory ($n_B$ in table 2). In this case, various measurements were performed at 1 mA / 1 T and 1 mA / 4.9 T, with the average and deviation values presented in table 2. Free carrier concentrations from infrared reflectometry are also shown in table 2 ($n_{IR}$) using the method already described in the body of the paper.

For the investigation of the Ge atom concentration we used a ToF-SIMS 5 system (ION-TOF GmbH) with a Bi-cluster primary ion gun and a Cs-sputter gun. All depth profiles were measured in the negative ion mode with 25 keV Bi⁺ as primary ion species and 1 keV Cs⁺ ions for the sputter process. The analysis area was 75×75 µm² and was placed in the center of the 150×150 µm² sputter area. The rastered area was 128×128 pixels with 10 shots/pixel. The measurements were performed in the non-interlaced mode with 1.5 s sputter time and 0/5 s low energy electron flooding per cycle for charge compensation. Four profiles were measured per sample and averaged to reduce the error.

For the quantification of the Ge concentration we used the method of relative sensitivity factors (RSFs) and therefore three reference samples with Ge concentrations 1×10¹⁹ cm⁻³, 5×10¹⁹ cm⁻³, and 1×10²⁰ cm⁻³, which were grown by MOVPE. The RSF was determined by:

$$C_x = RSF_x \frac{I_x}{I_R} \qquad (5)$$

and a linear regression using the depth profiles of the reference samples in a range of 100–600 nm (inside the film). In the equation, $C_x$ is the concentration of the species $x$, $I_x$ is the intensity of the species $x$, and $I_R$ is the intensity of the reference species $R$. Therefore, $C_x$ was plotted against $I_x/I_R$ and the $RSF$ was extracted from the slope of a linear fit.

We determined the RSF for different combinations of atomic mass/atomic number signals. It was observed that the combination of the ⁷¹Ga-($I_R$) and the combined (sum) signal ($I_x$) of Ge⁻, $^{70}$Ge⁻, $^{72}$Ge⁻, GeN⁻ showed the lowest error (0.1%). The RSF is often expressed in relation to the isotopic abundance and one atomic species - we determined a value of $5.5 \times 10^{20}$ cm$^{-3}$. Table 2 shows the Ge concentration extracted from ToF-SIMS. However, for values higher than mid-$10^{19}$ cm$^{-3}$, the quantification should be taken cautiously, due to the expected sputter yield variations owing to matrix effects as doping approaches alloy concentrations.

To validate the ToF-SIMS measurements, additional Ge SIMS profiles of sample E were performed at an independent company providing material analysis service (see result with its error bar in table 2). SIMS measurements proceeded under Cs+ ion sputtering and negative ion detection. During the analysis, we followed three isotopes of Ge ($^{73}$Ge, $^{74}$Ge, $^{76}$Ge). For quantification, we used a reference consisting of Ge-implanted GaN, which was measured in the same run. The three isotopes lead to the same Ge doping level, so we can conclude that no mass interference occurred during the analysis.

The Ge atom concentration in the most heavily doped sample (I) was additionally studied by Atom Probe Tomography (APT) in a CAMECA Flextap system, operated in laser pulsing mode with an ultraviolet laser at a temperature of 40 K. APT is based on the sequential field effect evaporation of individual atoms located at the surface of a needle-shaped sample. The chemical nature of the evaporated atoms is obtained from time-of-flight mass spectrometry: Each peak in the mass spectrum corresponds to a mass over charge ratio of a given ion; therefore, several peaks are related to a given chemical element depending on the number of isotopes and on the charge state of the ion. After elemental identification of all peaks, the elemental composition of the material is simply computed from the proportion of atoms of each species. The mass spectrum of sample E shows the presence of the various Ge isotopes in agreement with their natural abundance.

After quantification, $(9.4\pm0.5)\times10^{20}$ cm$^{-3}$ has been obtained for the concentration of Ge. The large error bar is due to the fact that the peaks attributed to Ge ions are very close to those of Ga.

The validity of the free carrier measurements performed using Hall techniques and infrared spectroscopy and the Ge concentration estimation performed using SIMS techniques and APT has been assessed here. All free carrier measurements are found to be consistent with each other within the given error bars (except for a deviation of a factor of two in infrared spectroscopy measurement of the highest doped sample, I). The Ge concentration estimated by the two SIMS measurement are also consistent with each other, but they are lower than the free carrier concentration obtained from Hall measurements. On the contrary, APT measurements puts the Ge concentration of the highest doped sample I slightly higher than the Hall measurement. These deviations are not related to the inhomogeneity of the samples.

# Tables

**Table 1.** Various parameters of the samples under study: Ge cell temperature ($T_{Ge}$), FWHM of the x-ray rocking curve ($\Delta\Omega$), room-temperature carrier concentration ($n$) and resistivity ($\rho$), strain along the (0001) axis ($\varepsilon_{zz}$), band gap ($E_G$) and Urbach's tail energy ($E_{Urb}$) extracted from transmission measurements at room temperature.

| Sample | $T_{Ge}$ (°C) | $\Delta\Omega$ (arcsec) | $n$ (cm$^{-3}$) | $\rho$ ($\Omega$cm) | $\varepsilon_{zz}$ (%) | $E_G$ at 300 K (eV) | $E_{Urb}$ at 300 K (meV) |
|---|---|---|---|---|---|---|---|
| A | 720 | 133 | 7.8×10$^{17}$ | 8.1×10$^{-2}$ | 0.25±0.02 | 3.432 | 52.2 |
| B | 760 | 392 | 1.0×10$^{18}$ | 1.71×10$^{-1}$ | 0.211±0.003 | 3.436 | 53.3 |
| C | 800 | 120 | 2.4×10$^{18}$ | 2.61×10$^{-2}$ | 0.084±0.008 | 3.42 | 58.9 |
| D | 850 | 60 | 1.3×10$^{19}$ | 6.03×10$^{-3}$ | 0.166±0.003 | 3.43 | 66.0 |
| E | 875 | 100 | 3.1×10$^{19}$ | 3.27×10$^{-3}$ | 0.207±0.008 | 3.438 | 72.4 |
| F | 900 | 287 | 6.8×10$^{19}$ | 6.00×10$^{-3}$ | 0.29±0.01 | 3.484 | 94.9 |
| G | 925 | 115 | 1.5×10$^{20}$ | 8.75×10$^{-4}$ | 0.244±0.003 | 3.522 | 117 |
| H | 950 | 296 | 2.6×10$^{20}$ | 5.51×10$^{-4}$ | 0.242±0.003 | 3.58 | 132 |
| I | 1000 | 193 | 6.7×10$^{20}$ | 6.90×10$^{-4}$ | 0.317±0.003 | 3.671 | 198 |
| X | n.i.d. | 237 | -- | -- | 0.236±0.003 | 3.443 | 46.9 |

**Table 2.** Measurements of carrier concentration and Ge atom density performed by various techniques: $n_A$, and $n_B$ are Hall effect measurements performed in different systems, $n_{IR}$ is the carrier concentration extracted from infrared reflectivity measurements, and [Ge]$_{ToF-SIMS}$, [Ge]$_{SIMS}$, and [Ge]$_{APT}$ are Ge atom concentrations measured by ToF-SIMS, SIMS and APT.

| Sample | $n_A$ (×10$^{18}$ cm$^{-3}$) | $n_B$ (×10$^{18}$ cm$^{-3}$) | $n_{IR}$ (×10$^{18}$ cm$^{-3}$) | [Ge]$_{ToF-SIMS}$ (×10$^{18}$ cm$^{-3}$) | [Ge]$_{SIMS}$ (×10$^{18}$ cm$^{-3}$) | [Ge]$_{APT}$ (×10$^{18}$ cm$^{-3}$) |
|---|---|---|---|---|---|---|
| A | 0.78±0.07 | | | 0.25±0.06 | | |
| B | 1.0±0.1 | | | 0.47±0.09 | | |
| C | 2.4±0.2 | | | 1.4±0.3 | | |
| D | 13±1 | 11±3 | | 6.3±1.4 | | |
| E | 31±3 | | | 18±3 | 14±3 | |
| F | 68±7 | | 80±7 | 22±4 | | |
| G | 150±20 | | 170±40 | 49±11 | | |
| H | 260±30 | | 260±10 | 93±20 | | |
| I | 670±70 | | 320±10 | 260±60 | | 940±50 |

# Figure captions

**Figure 1.** RHEED intensity oscillations during the Ga desorption after the growth of n.i.d. GaN and GaN:Ge growth at Ge cell temperatures $T_{Ge}$ = 800°C, 900°C, 950°C and 1000°C. The growth was interrupted at time $t = 0$ by shuttering the Ge, Ga and N cells simultaneously.

**Figure 2.** AFM images of samples X, F and I. (Top: 1×1 µm² surface: z range 0–5 nm. Down: 5×5 µm² surface: z range 0–9 nm).

**Figure 3.** (a) Evolution of carrier concentration with the Ge cell temperature. (b) Comparison of carrier concentration and Ge ToF-SIMS signal showing a linear relationship. (c) ToF-SIMS profiles of GaN:Ge sample H, and (d) n.i.d GaN sample X.

**Figure 4.** Infrared reflectance measured in samples F, G, H and I for p-polarized light at an angle of incidence of 70°. The measurements are corrected by the reflectance of a gold mirror, normalized, and vertically shifted for clarity. The features associated to the free carrier plasma are marked by arrows. They were differentiated from interference patterns by comparing measurements at various angles of incidence.

**Figure 5.** Variation of (a) free carrier concentration and (b) resistivity with temperature for various GaN:Ge samples. Dashed lines in (a) correspond to exponential fits leading to the activation energy ($E_a$) values indicated in the figure. Inset: Variation of $E_a$ as a function of the carrier concentration. The solid line is a fit to $E_a = E_I - \alpha(n)^{1/3}$.

**Figure 6.** Photoluminescence spectra of various samples at (a) 5 K and (b) 300 K. The spectra are normalized and vertically shifted for clarity.

**Figure 7.** Variation of the band gap energy obtained from Tauc's plot as a function of the carrier concentration. The dashed line corresponds to the calculation of $\Delta E_{BGR}$ shifted by the band gap energy of the n.i.d sample X, and the solid line corresponds to the calculation of $\Delta E_{BGR} + \Delta E_{BME}$ shifted by the band gap of sample X. Inset: Tauc's plot of selected samples.

**Figure 1**

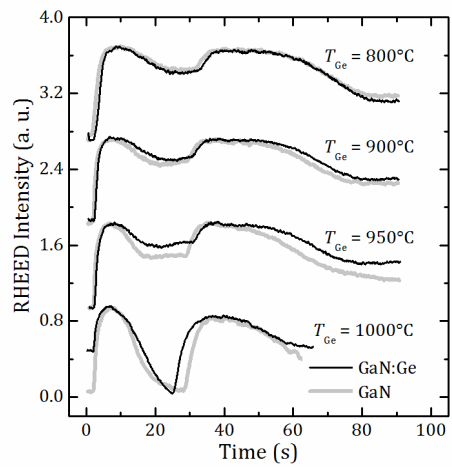

**Figure 2**

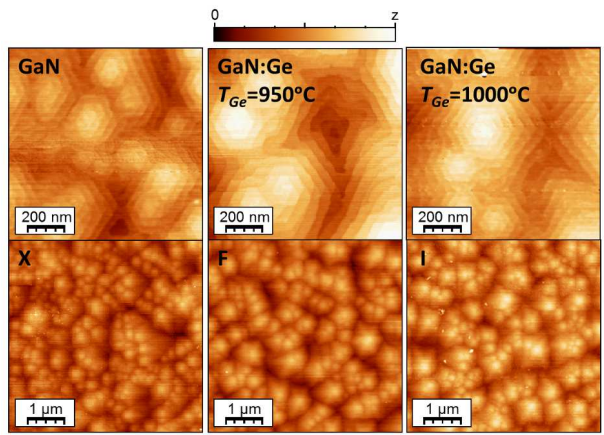

**Figure 3**

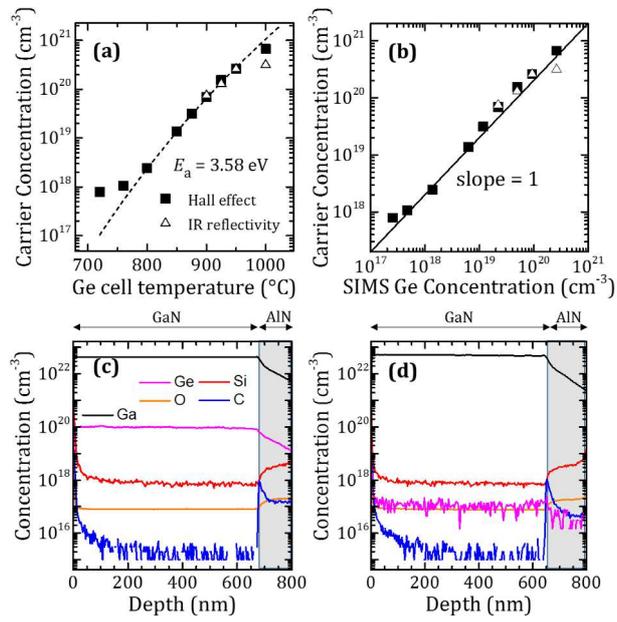

**Figure 4**

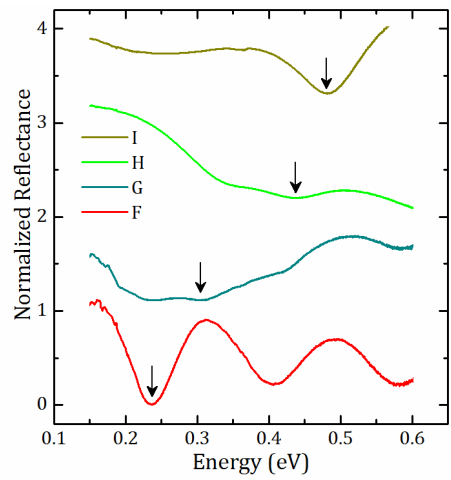

**Figure 5**

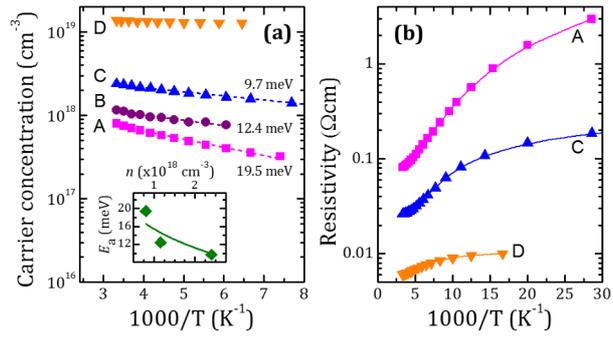

**Figure 6**

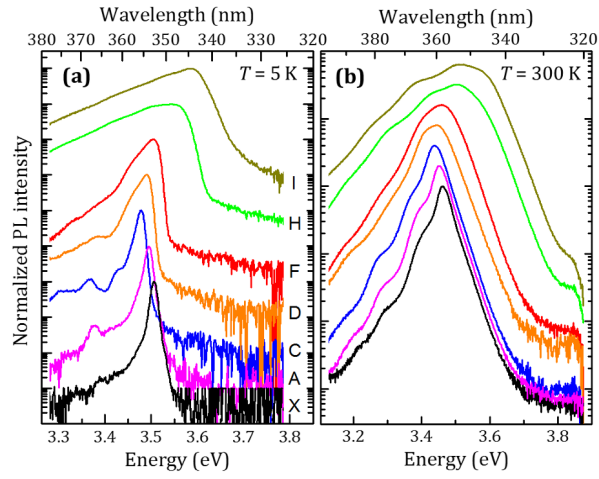

**Figure 7**

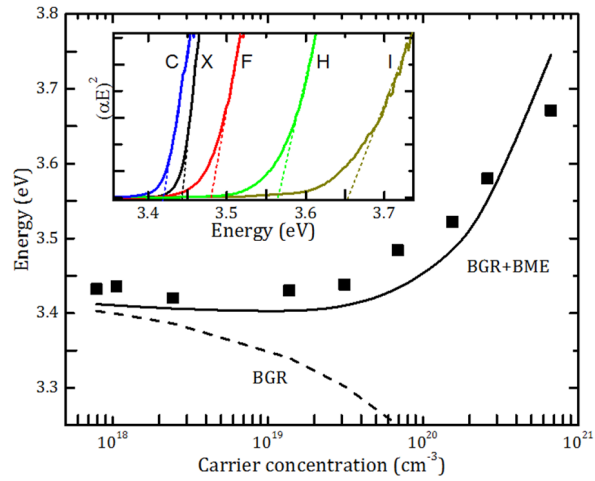